# Energy from food

David W.L. Hukins

Biodynamics Laboratory, Department of Surgery and Cancer, Imperial College London, Charing Cross Hospital Campus, London W6 8RF, UK

Department of Mechanical, Materials and Manufacturing Engineering, University of Nottingham, University Park, Nottingham NG7 2RD, UK

dhukins@imperial.ac.uk

Expressing the energy content of food as the heat energy released by its combustion is potentially misleading. Food is used to produce adenosine triphosphate (ATP). The free energy of conversion of ATP into adenosine diphosphate is used directly for processes such as muscular contraction, without the need for intermediate heat production. The number of ATP molecules produced depends on the task being performed and the fitness level of the person performing the task, since both affect the extent to which aerobic and anaerobic respiration are involved. The digestion and metabolism of soluble carbohydrates, complex carbohydrates and fats requires production of different enzymes. The information required to assess whether this affects the net ATP production by these different types of food does not appear to be available.

1.  Introduction

The purpose of this paper is to suggest that the way in which the energy content of food is measured is responsible for misleading ideas about the way in which energy from food is available to the human body. The energy content of food is measured by the heat energy released by its combustion [1]. It is expressed in kilo-joules or in the non-SI unit the kilo-calorie (often misleadingly called the "calorie" in the context of nutrition), which is specifically a measure of heat energy. Specifically, the kilo-calorie is the heat energy required to raise the temperature of 1 kg of water by 1°C. What is being measured is the heat generated by combustion. The concept of "counting calories" in which the "calorie content" of food is used when the body does work has been questioned previously using endocrinological evidence [2]. The current paper also questions this concept but is based on a consideration of the ways in which our bodies use energy.

2.  Efficiency of energy use

Heat energy is not the same as the energy that the body can obtain from food because the body is not a heat engine. Superficially, the body resembles a heat engine in which the carbon content of food is oxidised to carbon dioxide and the hydrogen content to water. These are the same products that occur in the exhaust of an efficient heat engine. The most efficient heat engine would be based on the Carnot cycle [3]. Since the body operates at a temperature of 37°C (310 K), its efficiency at an ambient temperature of 20°C (293 K) would be only 0.055. In an environment at 37°C, its efficiency would be zero, i.e. the total energy content of the food would be wasted and life could not be sustained. Heat can be considered as waste energy because it is the kinetic energy of molecules moving at random so they cannot do useful work.

The body contains molecular machines in which molecules do not move at random but instead make specific interactions to perform required tasks. The contraction of muscle provides an example [4]. A chemical reaction occurs between adenosine triphosphate (ATP), a myosin molecule in a thick muscle filament and an actin molecule in a thin filament. This chemical reaction causes a change in molecular conformation that pulls the thin filaments further between the thick filaments, so that the muscle contracts. In this process, ATP is converted to ADP and a phosphate ion. The mechanical work done by the muscle is derived directly from the free energy released by conversion of ATP to ADP; no heat production is involved. In contrast, a heat engine converts fuel into heat (by combustion) and this heat (waste energy) is then used to do useful work; it is the intermediate production of heat that limits its efficiency.

3. Digestion and metabolism

The energy available from different foods for the molecular machines of the body need to be known instead of their heat of combustion. In principle, the useful energy available from food could differ between soluble and complex carbohydrates, between carbohydrates and fats, for the reasons explained below. Further details on the metabolic processes described below are given, for example, in reference [5]. Useful energy is derived from the hydrolysis of a purine triphosphate (usually ATP) into the corresponding diphosphate. Therefore, the useful energy is the net energy that is available to synthesise ATP. For example, glucose is soluble in water and so requires no digestion for its absorption. In aerobic respiration, each glucose molecule absorbed can be used to convert 36 ADP molecules into ATP in a series of chemical reactions known as the Krebs or citric acid cycle, which involves production of acetyl coenzyme A (acetyl co A). Energy will be used in the production of the enzymes that catalyse each step of the process. Complex carbohydrates need to be digested to soluble sugars to be absorbed; this process involves expenditure of energy for the synthesis of digestive enzymes. Fats are digested by different chemical reactions to give fatty acids which are converted to acetyl co A by a different route than glucose; this acetyl co A then enters the Krebs cycle but, because different chemical reactions were involved previously, the energy used will be different.

The net energy from food is the energy available in the ATP molecules that are subsequently produced less the energy used in digestion and metabolism. In principle, the energy used will be different for soluble carbohydrates (like glucose), complex carbohydrates and fats. The question that then arises is whether the same quantities of digestive and other enzymes are produced irrespective of the proportion of these different food types in the diet. If the composition of the diet does not influence the production of the different types of enzymes, then it makes no difference whether energy is obtained from soluble carbohydrates, complex carbohydrates or fats. However, if the body responds to the composition of the diet by making more enzymes for carbohydrate or fat metabolism, then the useful energy will be different for the different food types. The extent of this difference and the energy differences involved would need to be measured to determine whether any such difference were appreciable. The information required to assess the useful energy from different foods, in this way, does not appear to be available in the literature.

4. Physical activity and physical fitness

Some forms of physical activity will use oxygen, required in aerobic respiration, more rapidly than it can be supplied. The effect of the type of physical activity can best be illustrated by a sprinter whose muscles contain insufficient oxygen to produce ATP by aerobic respiration [6]. At low oxygen concentrations, ATP can also be produced from glucose by anaerobic respiration (which does not require oxygen). However, in anaerobic respiration, one glucose molecule can convert only 2 ADP molecules into ATP, as compared with 36 converted by aerobic respiration [5]. So, the sprinter only obtains 1/18th of the energy from a glucose molecule that would be obtained by somebody going for a leisurely walk. In effect, the useful energy content food then depends on the type of physical activity that the body performs.

Finally, fitness levels influence how much energy can be obtained from food; training can reduce the need to produce ATP by anaerobic respiration [7].

5. Conclusions

Expressing the energy content of food as the heat generated by combustion is potentially misleading. The body uses food to produce ATP; the free energy generated by the hydrolysis of ATP, generating ADP and phosphate ions, is then used directly by the body without the need for any heat production. As an example, consider the energy content of glucose. For most activities, a single glucose molecule is used to produce 36 ATP molecules in aerobic respiration, a process that requires oxygen. However, in some activities, the muscles use oxygen more rapidly than it can be supplied; a single glucose molecule can then be used to produce only 2 ATP molecules by the process of anaerobic respiration. Therefore, the useful energy content of glucose depends on the nature of the tasks that the body needs to perform. Since training affects the need for anaerobic respiration, the energy obtained from one glucose molecule, in preforming a given task, will vary between different people.

The chemical reaction involved in processing soluble carbohydrates, complex carbohydrates and fats are different. Each of these reactions is catalysed by specific enzymes and energy is required to produce these enzymes. There appears to be insufficient information available to assess whether the energy involved in the production of enzymes affects the energy available from these different food types.

The practice of "counting calories" to determine the quantity of energy that can be obtained from food is potentially misleading for the reasons described above. Unfortunately, it would be very difficult to recommend an alternative simple rule for providing advice on how much energy can be obtained from different types of food.

**Acknowledgements**

I thank Professors Alison McGregor and Donal McNally for collaborative research on whole-body biomechanics which led to the ideas in this paper and Jean-Michel Desmarais for many interesting discussions on nutrition. I am grateful to Melanie Hargreaves (Registered Dietician), for comments on the first draft, and to Ben Hukins and Tom Hukins for help.